# Universal whirling magnetic orders in non-Heisenberg Tsai-type quasicrystal approximants

Farid Labib[1*], Kazuhiro Nawa[2*], Yusuke Nambu[3,4,5], Hiroyuki Takakura[6], Yoichi Ikeda[3], Kazuhiko Deguchi[7], Masato Matsuura[8], Asuka Ishikawa[1], Ryoichi Kajimoto[9], Kazuhiko Ikeuchi[10], Taku J. Sato[2] and Ryuji Tamura[11]

[1] *Research Institute of Science and Technology, Tokyo University of Science, Tokyo 125-8585, Japan*

[2] *Institute of Multidisciplinary Research for Advanced Materials (IMRAM), Tohoku University, Sendai 980-8577, Japan*

[3] *Institute for Materials Research, Tohoku University, Sendai, Miyagi 980-8577, Japan*

[4] *Organization for Advanced Studies, Tohoku University, Sendai, Miyagi 980-8577, Japan*

[5] *FOREST, Japan Science and Technology Agency, Kawaguchi, Saitama 332-0012, Japan*

[6] *Division of Applied Physics, Faculty of Engineering, Hokkaido University, Sapporo 060-8628, Japan*

[7] *Department of Physics, Graduate School of Science, Nagoya University, Nagoya 464-8602, Japan*

[8] *Neutron Science and Technology Center, Comprehensive Research Organization for Science and Society (CROSS), Tokai, Ibaraki 319-1106, Japan*

[9] *Materials and Life Science Division, J-PARC Center, Japan Atomic Energy Agency, Tokai, Ibaraki, 319-1195, Japan*

[10] *Institute of Materials Structure Science High Energy Accelerator Research Organization (KEK), 319-1106, Japan*

[11] *Department of Materials Science and Technology, Tokyo University of Science, Tokyo 125-8585, Japan*

Magnetic orders of non-Heisenberg Tsai-type 1/1 approximant crystals (ACs) in the Au-Ga-Dy system were studied through bulk magnetization, neutron diffraction, and inelastic neutron scattering techniques. The results uncovered noncoplanar, ferromagnetic (FM) and antiferromagnetic (AFM) spin configurations whirling along [111] crystallographic axis, which is analogous to those observed in the Tb- and Ho-contained counterparts. The crystal electric field excitations similar to those in the Tb-based counterpart are also observed indicating the strong Ising-like magnetic anisotropy. These comprehensive experiments and analyses have revealed the existence of a universal mechanism that stabilizes noncoplanar FM and AFM structures in non-Heisenberg Tsai-type ACs, independent of the rare-earth species (Tb, Dy, Ho); FM intra-cluster interactions and strong Ising-like anisotropy.

## I. INTRODUCTION

Quasicrystals (QCs), first discovered in 1984 [1], represent a fascinating class of solid-state materials possessing long-range order without periodicity in their atomic structure. Their atomic structure [as of Tsai-type [2]] comprise rare-earth (RE) elements having well-localized spins distributed on the vertices of icosahedron, as one of the inner shells of a multi-shell polyhedron, also called a rhombic triacontahedron (RTH) cluster [see Fig. 1(a)]. Whether aperiodic structures of icosahedral quasicrystals (iQCs) could host long-range magnetic orders has been a debating issue for decades until recently that ferromagnetic (FM) [3,4] and antiferromagnetic (AFM) [5] orders were discovered in the real iQCs.

Approximant crystals (AC), on the other hand, are closely related phases to QCs sharing the same RTH motif in their atomic structure but with periodic arrangement, as shown in Fig. 1(b) for the Tsai-type 1/1 AC. Owing to their structural similarity, ACs provide an ideal platform for studying magnetism in QCs, a subject that is far from being well understood. Through detailed investigations of magnetism in ACs in various alloy systems, it has been shown that reducing their electron-per-atom (*e/a*) via elemental substitution leads to emergence of long-range magnetic orders [6–9]. Neutron diffraction experiments on several Tb-containing FM and AFM 1/1 ACs have revealed a unique non-coplanar whirling magnetic structure [10–15]. Similar structure has also been reported in Ho-containing FM 1/1 AC elsewhere [16].

Later, theoretical studies performed on a realistic atomic structure of Tb- and Dy-based Tsai-type 1/1 ACs employing a point charge model [17,18] under the crystal electric field (CEF) effect not only identified the same whirling magnetic structure but also realized a wide spectrum of unexplored nontrivial magnetic textures on the RE icosahedron, each characterized by a unique topological charge number defined by $n = \Omega/(4\pi)$ with $\Omega$ being a solid angle spanned by the twelve localised spins on the RE icosahedron indicating high degeneracy of magnetic ground states on the icosahedral clusters. This feature together with competing interactions inherent to such complex structures lead to a natural expectation of diverse magnetic orders beyond what is explored thus far in non-Heisenberg Tsai-type compounds, wherein CEF effect is present. The implications of realizing these exotic states in reality extend beyond to future spintronic and magnetocaloric applications, as supported by the recent prediction of topological nodal magnons in Tsai-type 1/1 ACs [19]. However, without a well-defined framework or a clear understanding of the underlying mechanisms, such explorations could be costly and inefficient.

Driven by such a literature gap, in the present work, we uncover universal rule(s) governing the formation of noncoplanar whirling magnetic orders in non-Heisenberg Tsai-type ACs using wide set of experimental techniques including powder neutron diffraction and inelastic neutron scattering on the Au-Ga-Dy 1/1 ACs. The choice of Dy as the RE element in the present work is for several reasons: First, unlike extensively studied Tb-containing Tsai-type ACs [10–15], the magnetic properties of Dy-containing counterparts remain largely unexplored, partly due to high neutron capture cross-section of Dy that complicates neutron diffraction practice. Additionally, the distinct electron configuration of $Dy^{3+}$ ($4f^9$) compared to $Tb^{3+}$ ($4f^8$) is likely to influence the CEF potential and magnetic structure. Moreover, $Dy^{3+}$ has a larger magnetic effective moment than $Tb^{3+}$ (10.63 versus 9.72), which may modify the competing interactions within the icosahedral clusters, potentially leading to novel magnetic states. More importantly, investigating microscopic details of magnetic structures in Dy-based 1/1 ACs and comparing them with those of Tb- and Ho-based counterparts provides insight into the common mechanism driving whirling magnetic orders in non-Heisenberg Tsai-type ACs, which is a central objective of this work.

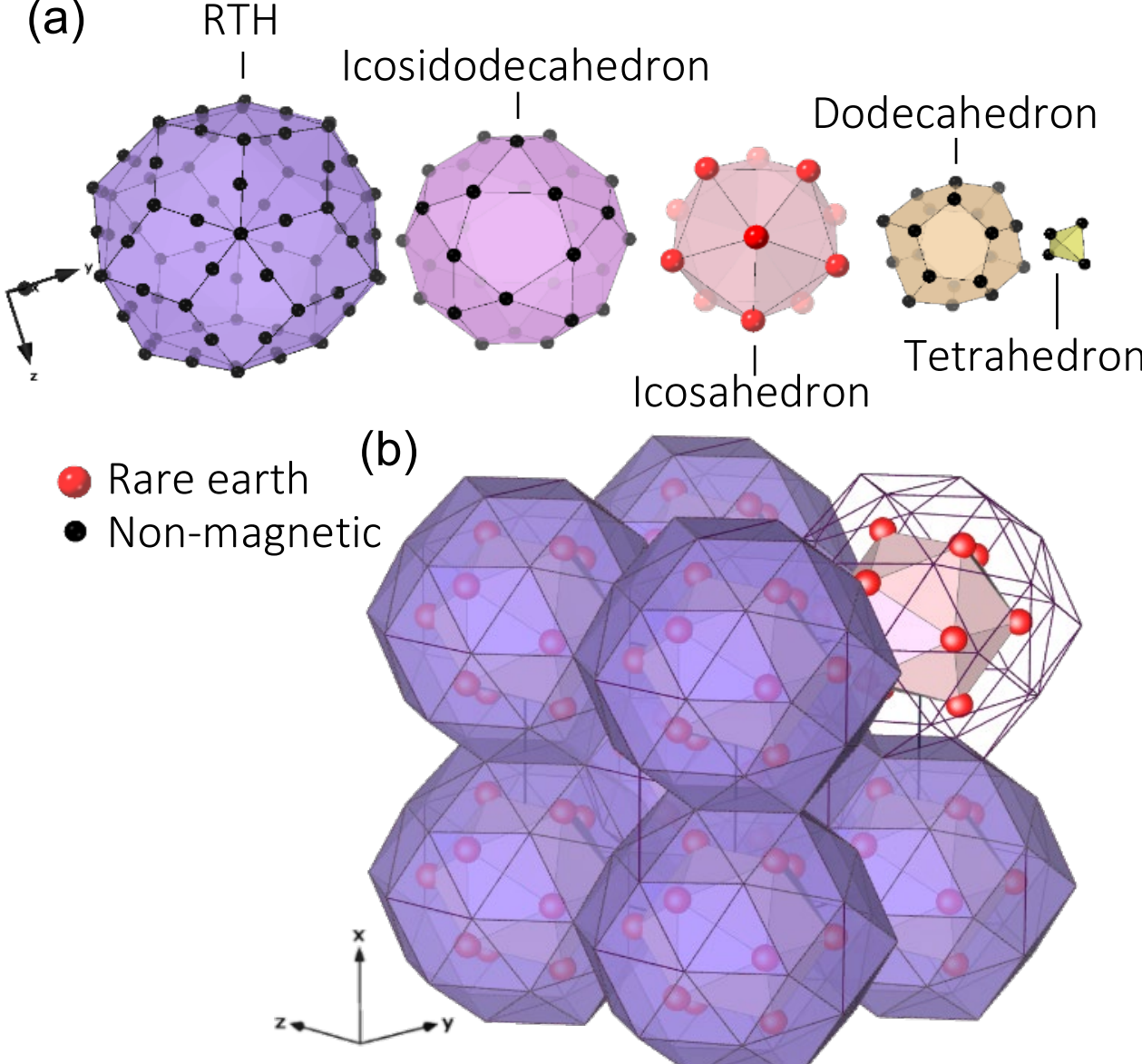

**Fig. 1.** (a) A shell structure of the Tsai-type compounds. From the outermost shell to the center: a rhombic triacontahedron (RTH) with 92 atoms sites, an icosidodecahedron (30 atomic sites), an icosahedron (12 atomic sites), a dodecahedron (20 atomic sites) and an innermost tetrahedron (4 atomic sites). (b) An arrangement of RTH clusters within a unit cell of a 1/1 AC and space group $Im\overline{3}$ .

## II. EXPERIMENT

Polycrystalline Au-Ga-Dy 1/1 ACs alloys varying in Au and Ga content but constant Dy of 14 at. % were synthesized employing an arc-melting technique followed by isothermal annealing at $T = 973$ K for 100 hours. The samples were examined for their dc magnetic susceptibility under zero-field-cooled (ZFC) and field-cooled (FC) conditions employing a superconducting quantum interference device magnetometer, Quantum Design: MPMS3 within $1.8 < T < 300$ K and in the external dc fields up to 7 T. Two of the synthesized 1/1 ACs with nominal compositions of $Au_{68}Ga_{18}Dy_{14}$ and $Au_{60}Ga_{26}Dy_{14}$ were selected for further investigation of their atomic and magnetic structures via single crystal X-ray diffraction (SCXRD) and powder neutron diffraction (PND) experiments. For indexing and integration of SCXRD peak intensities and for absorption correction, CrysAlisPRO software suite (Rigaku Oxford Diffraction, 2018) was implemented. The sample selection for PND experiment was based on their critical electron-per-atom (*e/a*) values that locate them inside a single FM or AFM regions within the universal *Curie-Weiss temperature* versus *e/a* framework unique to Tsai-type compounds [7]. For phase identification, Powder X-Ray diffraction (PXRD) was carried out using Rigaku SmartLab SE X-ray diffractometer with Cu-Kα radiation. For the PND experiment, a high-resolution powder diffractometer HERMES, installed at the Japan Research Reactor 3 (JRR-3) [20], is used. Neutrons with $\lambda = 1.34$ Å was selected using the Ge 551 reflections. For that purpose, 3.3 grams of $Au_{68}Ga_{18}Dy_{14}$ and $Au_{60}Ga_{26}Dy_{14}$ 1/1 ACs were packed in the Aluminum foil and shaped into a thin-walled cylinder with the diameter of 10 mm. The samples were then sealed in the Vanadium-Nickel can with He exchange gas. The sample can was set on the cold head of the closed cycle $^4$He refrigerator and cooled down to 2 and 15 K during the measurements. The absorption correction was numerically performed by assuming that the sample is shaped in an ideal thin-walled cylinder.

Inelastic neutron scattering experiments were performed using 4SEASONS installed in the Materials and Life Science Experimental Facility at Japan Proton Accelerator Research Complex (J-PARC) [21] to estimate CEF parameters. The main data were collected at the incident neutron energy of $E_i = 12.7$ meV using a Fermi chopper rotated at frequencies of 200 Hz. For that purpose, 4.1 grams of $Au_{65}Ga_{21}Dy_{14}$ 1/1 AC were packed in the Aluminum foil and shaped into a thin-walled cylinder with a diameter of 20 mm. The polycrystalline sample was cooled down to 6 K with He exchange gas using a top-loading closed-cycle $^4$He refrigerator. All the collected data were analyzed using the Utsusemi software suite [22] and a fortran code for a absorption correction [23].

## III. RESULTS

### III. I. Phase & structure characterization

Figure S1 in the Supplemental Material shows Powder X-Ray diffraction (PXRD) patterns of the prepared $Au_xGa_{86-x}Dy_{14}$ ($45 < x < 68$) 1/1 ACs alongside a calculated pattern derived from atomic structure of $Au_{68}Ga_{18}Dy_{14}$ 1/1 AC, refined in the present study (discussed later). The experimental positions and intensity distributions of the peaks align closely with the calculation, indicating mutual substitution of Au and Ga within the structure. Such mutual substitution results in a wide single-phase region in the chemical phase diagram of Au-Ga-Dy enabling one to tune the magnetic properties by solely altering their *e/a*, as explored elsewhere [13]. Among the synthesized ACs, two of them with nominal compositions of $Au_{68}Ga_{18}Dy_{14}$ and $Au_{60}Ga_{26}Dy_{14}$ 1/1 ACs are selected in the present work due to their respective AFM and FM ground states.

To gain insight into their atomic structure, single crystal X-ray diffraction (SCXRD) experiments were conducted on both samples. Figure 2 provides the refined atomic structure model of the $Au_{68}Ga_{18}Dy_{14}$ 1/1 AC, viewed along the [100] direction, while the structure model of the $Au_{60}Ga_{26}Dy_{14}$ 1/1 AC is provided separately in Fig. S2 of the Supplemental Material. The SCXRD data refinement was carried out using SHELXT [24] for initial models and SHELXL [25] for final refinement. Crystallographic data and refinement parameters of the SCXRD experiments as well as the final model including atomic coordinates, Wyckoff positions, site occupations, and equivalent isotropic displacement parameters ($U_{eq.}$) are listed in Table S(1,2) and Table S(3,4) of the Supplemental Material, respectively. In Figs. 2 and S2 (in Supplemental Material), Au, Ga, Dy atoms and unoccupied sites are represented by yellow, dark blue, red and white spheres, respectively. Clearly, in both 1/1 ACs, the primary building unit is the RTH cluster consisting of three concentric inner shells and a central disordered tetrahedron unit. The lattice parameter of the $Au_{68}Ga_{18}Dy_{14}$ 1/1 AC is 14.7313(1) Å being slightly larger than that of the $Au_{60}Ga_{26}Dy_{14}$ 1/1 AC reflecting a lattice expansion due to slightly higher Au content in the structure. Notably, the centers of the cube-shaped interstices connecting adjacent dodecahedra along 3-fold axes (Ga8 site) and 24 vertices of a rhombic triacontahedron (Ga7 site) are exclusively occupied by Ga. The next preferred site for Ga is 12 vertices of a dodecahedron, which are occupied by both Au and Ga atoms (Au/Ga6 site). Figure 3 depicts the closest neighboring environment around a $Dy^{3+}$ atom in the actual structural model of $Au_{68}Ga_{18}Dy_{14}$. The occupancy-weighted distance of the sites close to the $Dy^{3+}$ ions in this compound are 3.119 Å (Au4), 3.147 Å (Au2), and 3.168 Å (Au/Ga6).

It is important to note that Ga partially replaces Au sites in the structure by increasing its at. % in the nominal composition. In the case of Au60Ga22Dy14, the occupancy of Ga atoms increases at the vertices of a dodecahedron (Au/Ga6 site) and 24 vertices of a rhombic triacontahedron (Au/Ga4 site), as shown in Fig. S2. The occupancy-weighted distance of the sites close to the $Dy^{3+}$ ions in the $Au_{60}Ga_{22}Dy_{14}$ become 3.108 Å (Au/Ga6), 3.128 Å (Au2), and 3.141 Å (Au2). The increased Ga occupation, therefore, makes a single Au/Ga6 site closer to $Dy^{3+}$ ions, whereas the distance to a single Au/Ga4 site becomes larger (3.149 Å). Such variation in the local environment of Dy atoms would influence magnetic anisotropy via a CEF potential, as will be discussed in detail later. In the following sub-sections, magnetic properties of the $Au_{68}Ga_{18}Dy_{14}$ and $Au_{60}Ga_{22}Dy_{14}$ 1/1 ACs will be examined in detail.

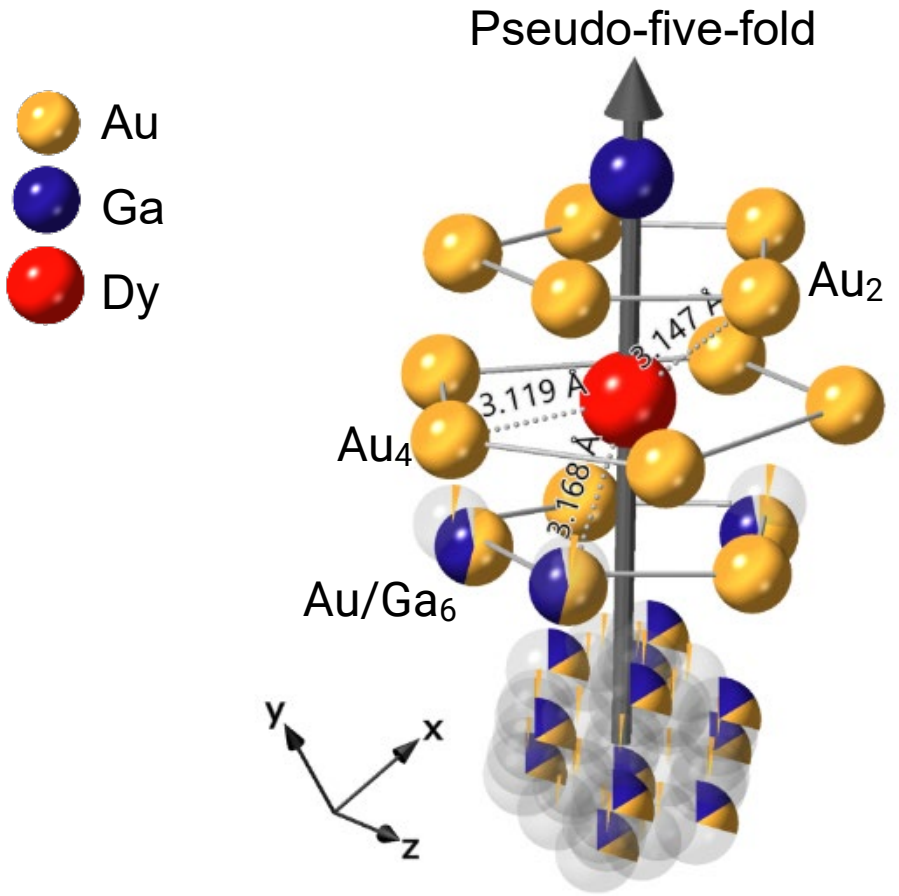

**Fig. 3.** The closest neighboring atoms around a $Dy^{3+}$ atom in the structural model of $Au_{68}Ga_{18}Dy_{14}$. The central arrow indicates a pseudo-five-fold axis.

## III. II. Magnetic properties

### III. II. i. Antiferromagnetism in Dy-contained 1/1 ACs

The $Au_{68}Ga_{18}Dy_{14}$ 1/1 AC is confirmed to be a single phase 1/1 AC with space group $Im\overline{3}$ and a lattice parameter of 14.7313(1) Å, as shown by the powder XRD pattern and refined atomic structure. Figure 4(a) presents the temperature dependence of the magnetic susceptibility exhibiting a distinct cusp at $T_N$ = 6.7 K in both the ZFC and FC magnetization curves, indicative of an AFM transition. In the paramagnetic region, the inverse magnetic susceptibility, shown in the inset of Fig. 4(a), displays a linear behavior following the Curie-Weiss law ($\chi(T) = \frac{N_A \mu_{eff}^2 \mu_B^2}{3k_B(T-\theta_w)} + \chi_0$), where, $N_A$, $\mu_{eff}$, $\mu_B$, $k_B$, $\theta_w$, and $\chi_0$ denote the Avogadro number, effective magnetic moment, Bohr magneton, Boltzmann constant, Curie-Weiss temperature, and the temperature-independent magnetic susceptibility, respectively. The linear fit of the data between 100 K and 300 K results in $\theta_w$ and $\mu_{eff}$ of +7.3±0.3 K and 10.67$\mu_B$, respectively. Figure 4(b) illustrates field dependence of magnetization (*M-H*) in the $Au_{68}Ga_{18}Dy_{14}$ 1/1 AC up to 7 T at 1.8 K (below $T_N$). The first derivative of the *M-H* curve in the upper-left inset of Fig. 4(b) evidences the appearance of a critical point at $H$ = 0.22 T corresponding to a metamagnetic transition under the applied magnetic field. A metamagnetic transition corresponds to a sudden spin reorientation within the AFM phase when a sufficiently strong external magnetic field is applied, leading to a forced FM or other magnetic phases. The narrow hysteresis observed in the magnified *M-H* curve in the bottom- right inset of Fig. 4(b) suggests the spin-reorientation transition to be of first order type. Such hysteresis, also seen between the FC and ZFC susceptibilities in Fig. 4(a), is due to the spontaneous magnetic moment being approximately 0.083$\mu_B$, as shown inside the bottom-right inset of Fig. 4(b). The hysteresis in the *M-H* curve has also been reported in other AFM 1/1 ACs including $Cd_6Tb$ [26], and $Au_{65}Ga_{21}Tb_{14}$ [10]. In addition, maximum magnetization of AFM $Au_{68}Ga_{18}Dy_{14}$ 1/1 AC approaches to ~ 6$\mu_B$/$Dy^{3+}$ at $T$ = 1.8 K and $H$ = 7 T (~ 70% of 10.00$\mu_B$/$Dy^{3+}$) indicating the presence of CEF-induced uniaxial anisotropy in the $Dy^{3+}$ spins.

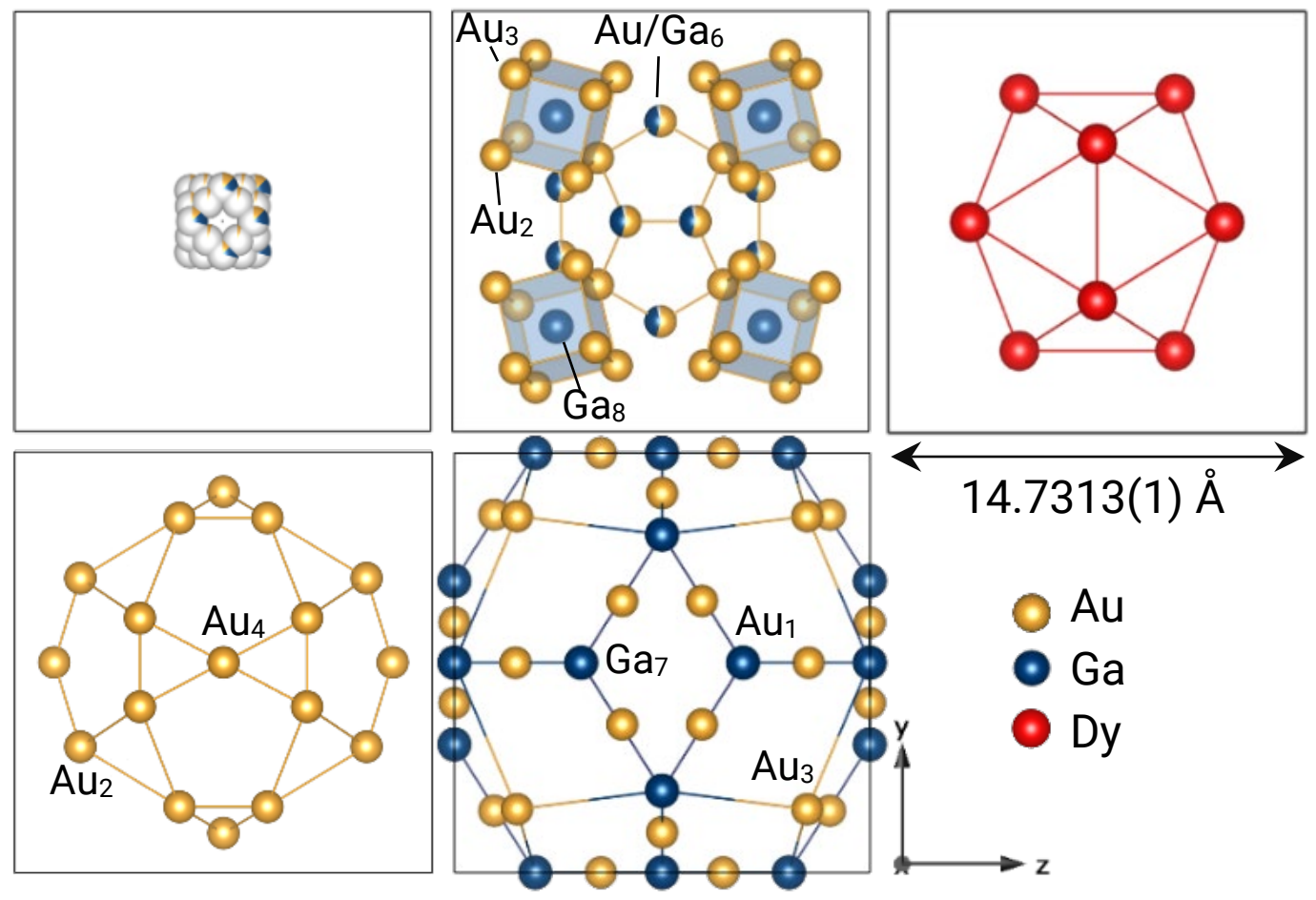

**Fig. 2.** The successive sequence of atomic shells of the $Au_{68}Ga_{18}Dy_{14}$ 1/1 AC; the inner most unit (top left corner) comprised of orientationally disordered tetrahedron. The vertices of dodecahedron (top middle) are occupied by Au and Ga while the centers of eight cube-shaped interstices connecting adjacent dodecahedra along 3-fold axes are occupied by Ga interstices (Ga8 site). The icosahedron (top right) and icosidodecahedron (bottom left) are exclusively occupied by Gd and Au, respectively. The vertices and mid-edge positions of the rhombic triacontahedron (RTH) in the bottom middle panel are occupied by Ga and Au atoms, respectively.

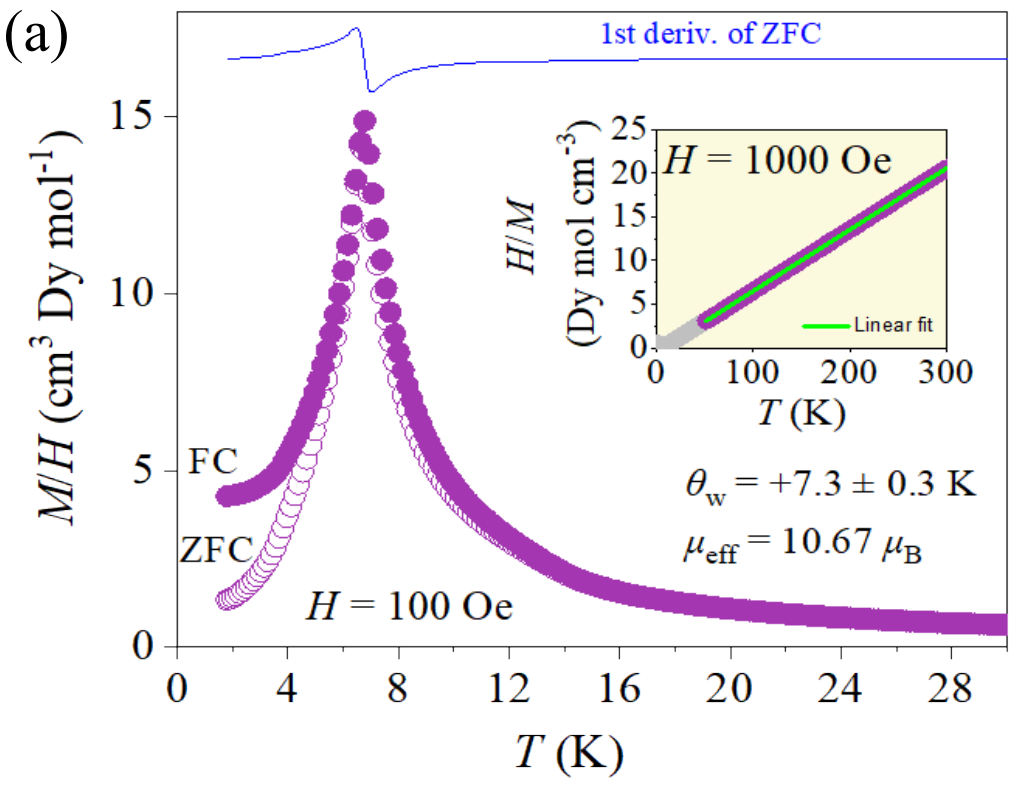

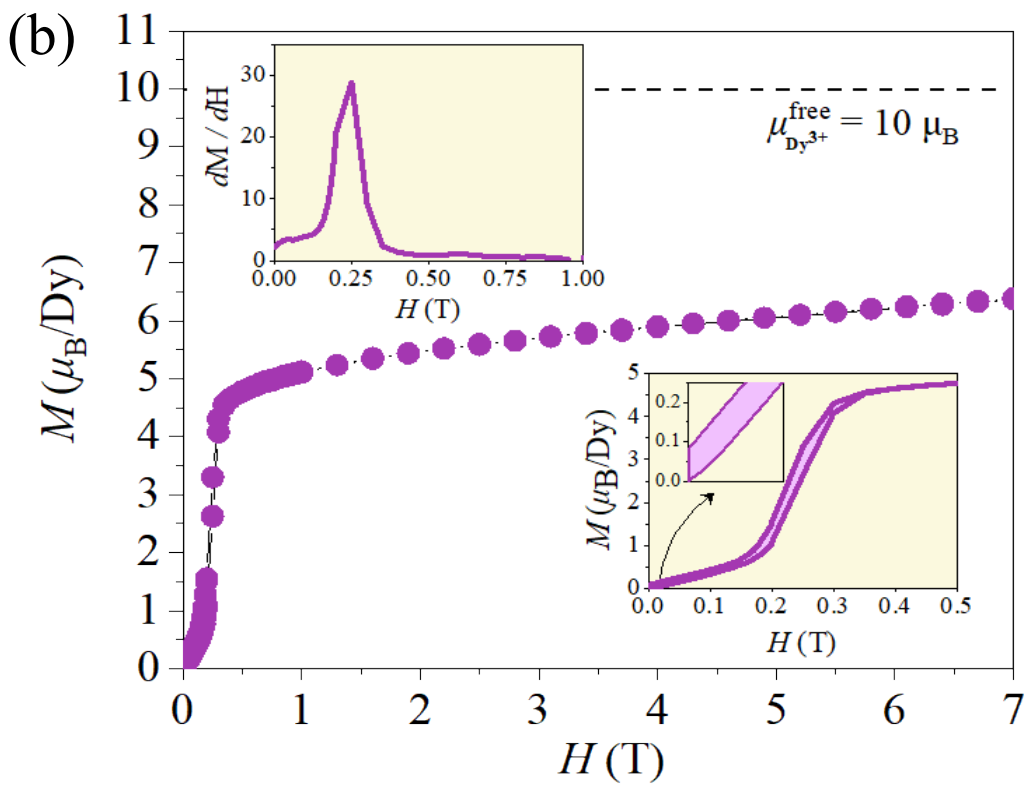

**Fig. 4. (a).** Low temperature magnetic susceptibility (*M/H*) of the $Au_{68}Ga_{18}Dy_{14}$ 1/1 AC as a function of temperature under FC and ZFC modes and $H$ = 100 Oe. The inset shows the inverse magnetic susceptibility (*H/M*) of the same compound. (b) Magnetic field ($H$) dependence magnetizations measured at 2 K up to $H\mu_0$ = 7 T. The upper-left inset in (b) shows first derivative of the *M-H* curve exhibiting a meta-magnetic transition at $H\mu_0$ = 0.22 T, while the bottom-right inset provides a magnified view if the *M-H* curve clearly evidencing a spontaneous magnetization.

Figure 5 (a) compares PND patterns for the $Au_{68}Ga_{18}Dy_{14}$ 1/1 AC at (a) $T$ = 15 K and (b) $T$ = 2 K. At 2 K, magnetic Bragg reflections appear at the indices *hkl* with $h + k + l = 2n + 1$ ($n$: an integer), which indicates the break of the BCC symmetry by the AFM order. The strongest magnetic Bragg reflection is 021 appearing at $2\theta$ = 11.0°. The intensity distribution of the magnetic reflections is quite similar to those of the whirling AFM order in the Tb-based 1/1 ACs [10–13] except that the intensity of the strongest magnetic reflections (021) is almost 1/3 of that of the strongest nuclear reflections in $Au_{68}Ga_{26}Dy_{14}$, while they are quite comparable in the Tb-based 1/1 ACs [13].

To confirm the crystal structures and solve the magnetic structures, the Rietveld refinement of the PND pattern is carried out following the approach discussed in refs. [10–13]. Here, the structural parameters obtained by SCXRD experiments are used as initial parameters to refine the crystal structures. The refined crystallographic data and structural parameters are listed in Tables S(5–7) of the Supplemental Material. For constructing magnetic structure, the magnetic representation theory [27] is applied, which justifies the magnetic representations of the magnetic moments to be decomposed as a linear combination of magnetic basis vectors within a single irreducible representation (IR) of the '*k*group' with $k$ = (1,1,1) associated with the crystallographic space group. The result of the decomposition and corresponding magnetic basis vectors (BVs) for all the IRs are obtained (see Table S(8,9) for the list of IRs and the atomic coordinates of the Dy1-12 sites, respectively). The fit at 15 K shows a good agreement with experimental curve using the structural model shown in Fig. 2 [see Fig. S3(a) and Table S6 in Supplementary Material for details]. A fit at 2 K (see Table S7) also yields a good agreement with experimental pattern when the magnetic structure belonging to IR2 is used for the refinement, as shown by a solid black line in Fig. 5(b). The refined magnetic structure of the $Au_{68}Ga_{18}Dy_{14}$ 1/1 AC along with that of the $Au_{65}Ga_{21}Tb_{14}$ 1/1 ACs are provided in Fig. 6(a) and (b), respectively, for the sake of comparison. The magnetic structures of the Dy- and Tb-contained AFM 1/1 ACs are similar being equivalent to that represented by a magnetic space group of $I_pm'3'$ (204.5, OG setting). Due to the presence of the threefold rotational symmetry, the spins form a whirling configuration around [111] crystallographic axis, therefore is denoted as "*whirling*" AFM order. In Fig. 6(a) and (b), one can see that the magnetic moments are fixed within the local mirror plane by symmetry. The canting angle between the spins and the pseudo-fivefold axis amounts to 75(1)° in the AFM $Au_{68}Ga_{18}Dy_{14}$ 1/1 AC, which is slightly smaller than 85(1)° in the Tb-based AFM 1/1 AC, shown in Fig. 6(b) [10]. Moreover, the magnitude of the magnetic moment in the AFM $Au_{68}Ga_{18}Dy_{14}$ 1/1 AC is estimated to be 5.58(12), which is also smaller than 7.86(14)$\mu_B$ found in AFM $Au_{65}Ga_{21}Tb_{14}$ 1/1 AC [10]. Table 1 compares moment size and canting angles in the Dy- and Tb-contained AFM 1/1 ACs.

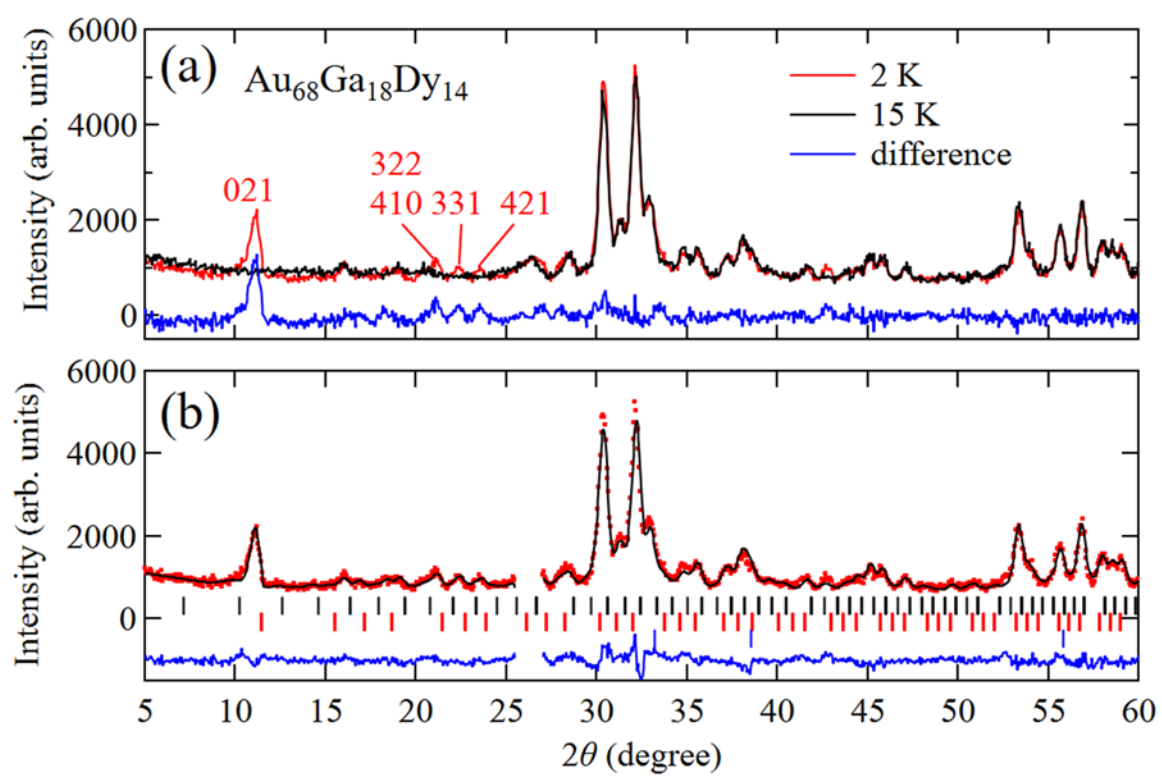

**Fig. 5.** (a) Powder neutron diffraction patterns of $Au_{68}Ga_{18}Dy_{14}$ collected at 2 (red) and 15 K (black). The blue curve indicates their difference. (b) Rietveld refinement of $Au_{68}Ga_{18}Dy_{14}$. Observed, calculated, and the difference between both intensities are shown by red dots, black, and blue curves, respectively. Nuclear and magnetic reflections for the antiferromagnetic order are indicated by the upper black and lower red vertical lines, respectively. Blue vertical lines represent reflections from the Al foil. The 2θ range between 25.5 and 27 degrees are excluded from the refinement to suppress the influence of unknown impurities.

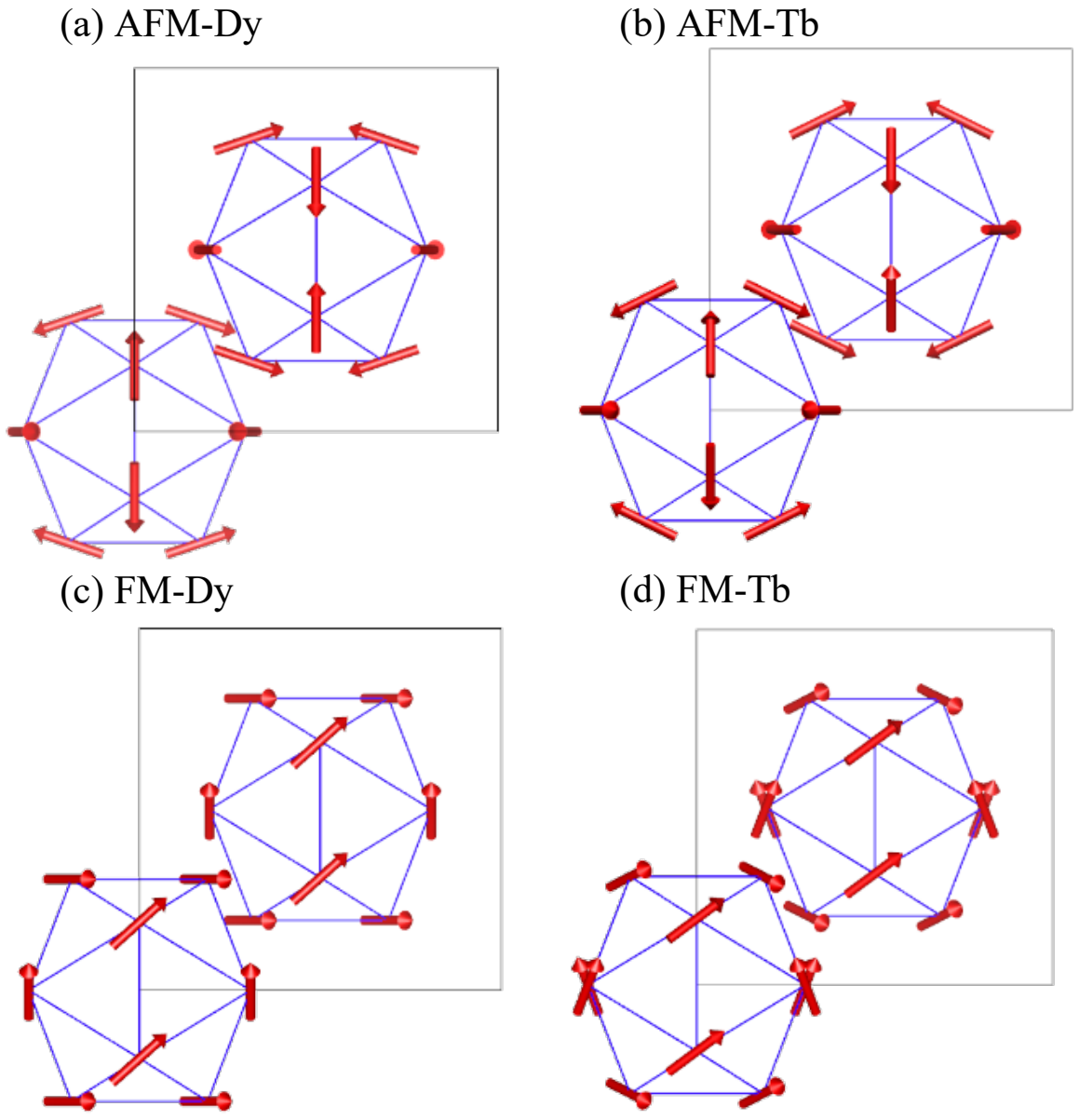

**Fig. 6.** Representation of magnetic structures on two adjacent icosahedron clusters in antiferromagnetic (a) $Au_{68}Ga_{18}Dy_{14}$, (b) $Au_{65}Ga_{21}Tb_{14}$ 1/1 ACs [10] and ferromagnetic (c) $Au_{60}Ga_{26}Dy_{14}$, and (d) $Au_{60}Ga_{26}Tb_{14}$ 1/1 ACs [13]. Red arrows indicate magnetic moments.

### III. II. ii. Ferromagnetism in Dy-contained 1/1 ACs

Figure 7(a) presents the temperature dependent dc magnetic susceptibility ($M/H$) of $Au_{60}Ga_{26}Dy_{14}$ 1/1 AC over a temperature range of 0 – 30 K measured under FC and ZFC conditions (indicated by filled and unfilled circles, respectively). The inset shows the inverse magnetic susceptibility ($H/M$) of the same compound from 1.8 to 300 K, which follows a linear trend consistent with the Curie–Weiss law. From the linear fit of the data between 100 K and 300 K, $\theta_w$ is estimated as +6.3 ± 0.2 K with $\chi_0$ approximating zero. The $\mu_{eff}$ is found to be 10.67$\mu_B$, close to the theoretical value for free $Dy^{3+}$ defined as $g_J(J(J+1))^{0.5}$ $\mu_B$ [28], indicating that the magnetic moments are localized on $Dy^{3+}$ ions.

At low temperatures, as shown in Fig. 7(a), there exists a sharp rise in the magnetic susceptibility below $Tc$ = 8.3 K, determined from the minimum of $d(M/H)/dT$). Such rise of magnetic susceptibility is due to the onset of spontaneous magnetization ($M_s$) following the relation $M_s(T) = M_0(-\epsilon)^\beta$, where $M_0$, $\varepsilon$, and $\beta$ are the critical amplitude, reduced temperature ($T$ – $T_C$)/ $T_C$ and the critical exponent, respectively.

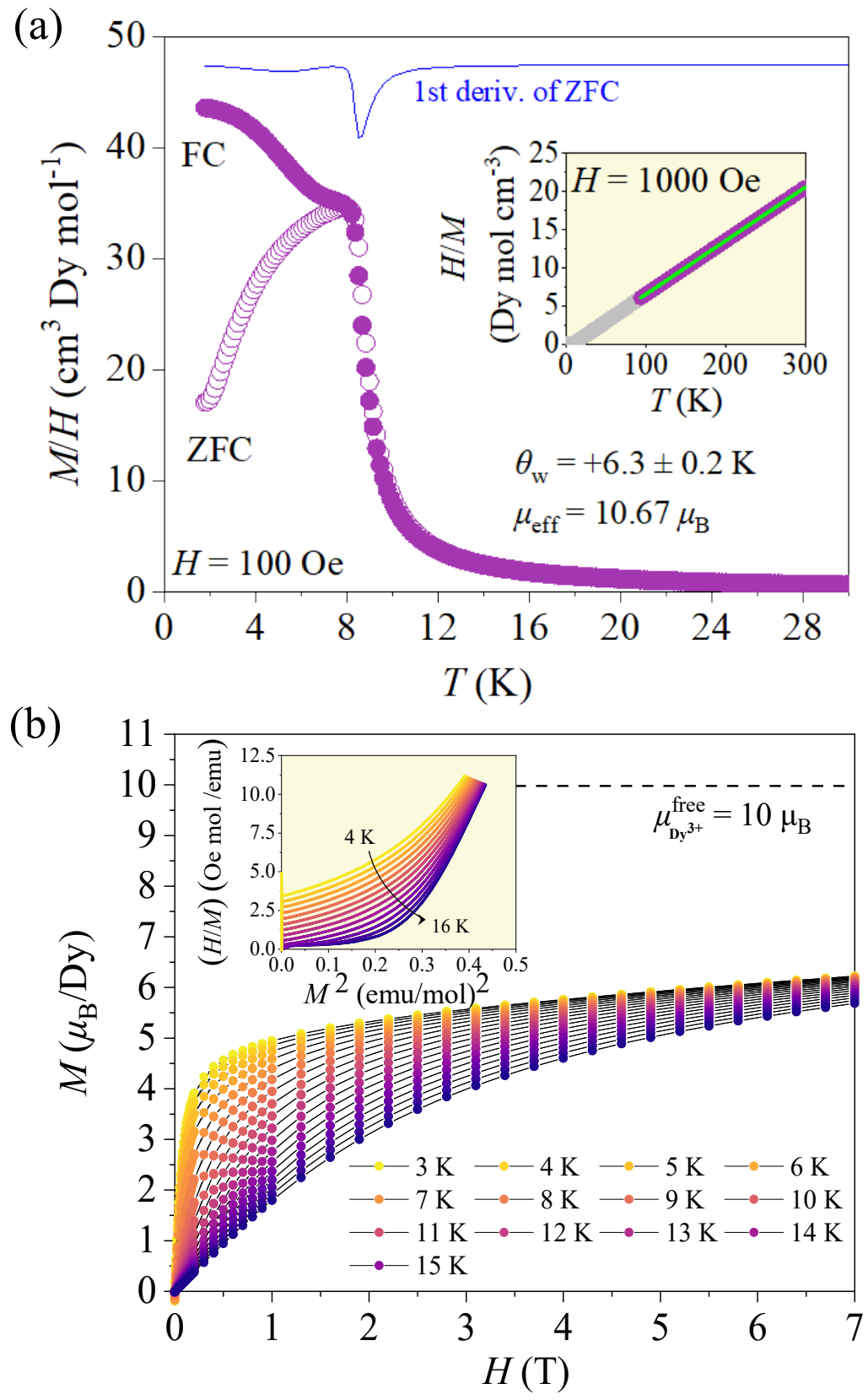

**Fig. 7.** (a) Low temperature magnetic susceptibility ($M/H$) of the $Au_{60}Ga_{26}Dy_{14}$ 1/1 AC as a function of temperature under FC and ZFC modes and $H$ = 100 Oe. The inset shows the inverse magnetic susceptibility ($H/M$) of the same compound. (b) Series of magnetic field ($H$) dependence magnetizations measured within a temperature range of 2 – 30 K up to $H$ = 7 T. In the inset of (b), Arrott plot ($H/M$ vs. $M^2$) of the same sample is provided.

Figure 7(b) presents isothermal magnetization curves versus magnetic field ($M$-$H$) for the $Au_{60}Ga_{26}Dy_{14}$ 1/1 AC across $T$ = 3 – 15 K. Clearly, below $T_C$ = 8.3 K, magnetization increases rapidly with the applied field reaching ~ 6$\mu_B$/$Dy^{3+}$ (~ 70 % of the full moment of a free $Dy^{3+}$ ion) at $T$ = 3 K and $H$ = 7 T. As the temperature increases, the curvature of magnetization becomes milder, indicating a transition towards paramagnetic region. The inset of Fig. 7(b) displays Arrott plots ($H/M$ vs. $M^2$) for the same compound exhibiting no sign of negative slope or inflection point

**Table 1.** The list of the magnitudes and the canting angles of magnetic moments with respect to the pseudo-fivefold axis.

| Compound | moment size ($\mu_B$) | canting angle (°) |
| --- | --- | --- |
| $Au_{60}Ga_{26}Dy_{14}$ (FM) – this work | 4.61(18) | 70(4) |
| $Au_{68}Ga_{18}Dy_{14}$ (AFM) – this work | 5.58(12) | 75(1) |
| $Au_{60}Ga_{26}Tb_{14}$ (FM) [13] | 5.58(23) | 75(2) |
| $Au_{65}Ga_{21}Tb_{14}$ (AFM) [10] | 7.86(14) | 85(1) |
| $Au_{61.1}Al_{25.3}Ho_{13.6}$ (FM) [16] | 9.219 (8) | – |

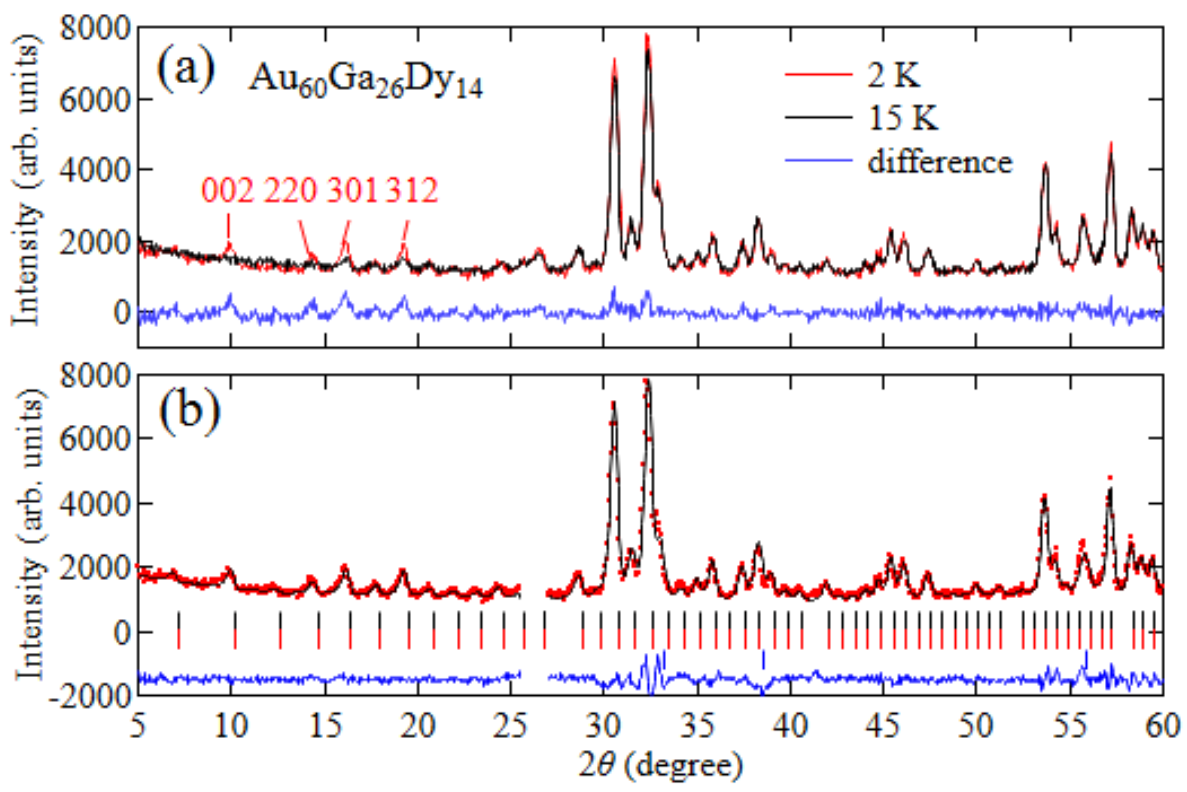

**Fig. 8.** (a) Powder neutron diffraction patterns of $Au_{60}Ga_{26}Dy_{14}$ collected at 2 (red) and 15 K (black). The blue curve indicates their difference. (b) Rietveld refinement of $Au_{60}Ga_{26}Dy_{14}$. Observed, calculated, and the difference between both intensities are shown by red dots, black, and blue curves, respectively. Nuclear and magnetic reflections for the antiferromagnetic order are indicated by the upper black and lower red vertical lines, respectively. Blue vertical lines represent reflections from the Al foil. The $2\theta$ range between 25.5 and 27 degrees are excluded from the refinement to suppress the influence of unknown impurities.

around $T_C$ indicating a second-order phase transitions, based on the Banerjee's criterion [29].

Figure 8(a) compares the PND patterns of the $Au_{60}Ga_{26}Dy_{14}$ 1/1 AC at $T = 2.0$ K and 15 K, corresponding to below and above $T_C = 8.3$ K, respectively. The magnetic reflections appear on top of the nuclear reflections with the indices *hkl* with $h + k + l = n$ (*n*: an integer) indicating that the BCC symmetry is preserved by the FM order. For instance, the strong magnetic reflection appears at $2\theta = 9.9^{\circ}$ (corresponding to 002). The intensity distribution of the magnetic reflections is similar to those of the noncoplanar FM order in the Tb-based 1/1 ACs reported elsewhere [8].

To confirm the crystal structure model, the Rietveld refinement of the PND patterns is carried out following the approach discussed earlier in sub-section 2.2.1. Similarly, the structural parameters derived from SCXRD experiments are used as initial parameters in refinement. The refined crystallographic data and structural parameters are listed in Tables S(5,8,9) of the Supplemental Material. The fit at 15 K shows good agreement with the experimental curve using the structural model provided in Fig. S2 of the Supplemental Material (see Fig. S3(b) in the Supplemental Material for the whole neutron powder diffraction pattern and Table S10 for the structural parameters). To find the magnetic structure, the magnetic modulation vector is selected as $k = (0,0,0)$ since the BCC symmetry is preserved by the magnetic order. A fit at 2 K, represented by a solid black line in Fig. 8(b) (see Table S11), also yields a good agreement with the experimental pattern when the magnetic structure belonging to IR8 (see Table S8 for a list of IRs) is used for the refinement.

In principle, it is not possible to determine the noncoplanar ferromagnetic structure in a cubic crystal structure since magnetic reflections induced by FM order are highly overlapped and their intensities are averaged out in powder diffraction patterns. Thus, to propose a possible magnetic structure, the FM phase is assumed to belong to the magnetic space group $R\overline{3}$, as reported in the other FM 1/1 ACs [14], leading to an inherent 3-fold symmetry along one of the diagonal directions. This magnetic space group results in two inequivalent magnetic atoms. In addition, it is assumed that the magnitudes of the magnetic moments across Dy sites are identical. In terms of magnetic representations of the Dy magnetic moments, the above assumption can be fulfilled by using basis vectors of IR8 under some restraints (see caption in Table S5 for details). From the Rietveld refinement using 3 adjustable parameters, two possible magnetic structures are derived. As compared in Table S12, both fits show reasonable agreement. The one has a small component and the other has a large component perpendicular to the local mirror plane. We deduce that the former is more likely, since the easy-axis direction expected from the moment direction is closer to that of $Au_{68}Ga_{18}Dy_{14}$.

The fit and the magnetic structure obtained from the refinements are depicted by solid black curve in Fig. 8(b) and Fig. 6(c), respectively. For comparison, the magnetic structure of the FM $Au_{60}Ga_{26}Tb_{14}$ 1/1 AC [13] is also provided in Fig. 6(d). Similar to the AFM samples, the magnitude of magnetic moment in the FM $Au_{60}Ga_{26}Dy_{14}$ 1/1 AC is smaller than the Tb-contained FM counterpart ($4.61(18)\mu_B$ versus $5.58(23)\mu_B$) despite the fact that magnetic effective moment of $Dy^{3+}$ is larger than that of the $Tb^{3+}$ (10.63 versus 9.72). Moreover, the canting angle against the pseudo five-fold axis, as compared in Table I, is about 5° deviated in the Dy-based FM sample compared to the Tb-based counterpart which may account for the reduced ordered moment size in the former. Deviation of the magnetic moments from the direction perpendicular to the pseudo fivefold axis has also been observed in the FM $Au_{61.1}Al_{25.3}Ho_{13.6}$ 1/1 AC [16]. Both Dy- and Tb-based FM 1/1 ACs exhibit out-of-mirror plane magnetic components with canting angles of ~ 48 –50°.

As summarized in Table I, the Dy-contained 1/1 ACs (either FM or AFM) showcase two clear features. First, the moment directions are almost the same as those of the Tb- and Ho-based ACs [10–13,16], indicating the strong dominance of uniaxial anisotropy of magnetic moments on the magnetic structures of the non-Heisenberg 1/1 ACs, regardless of RE type. The magnetic moments in the Dy- and Ho-contained ACs are, however, slightly away from the direction perpendicular to the pseudo fivefold rotation axis. Second, the ordered moment size in the Dy-contained samples is smaller than those in the Tb- and Ho-contained counterparts despite the larger magnetic effective moment of the $Dy^{3+}$ than the others. At the moment, parameters contributing to this behavior remain unclear. It is unequivocal, though, that chemical disorder around RE atoms plays a role and is even more crucial in the present Dy-contained samples.

### III. II. iii. Inelastic neutron scattering

The refined magnetic structures of the FM 1/1 ACs, either Dy- or Tb-based, suggests Ising-like character for the magnetic moment implying the existence of strong easy-axis anisotropy. Since the anisotropy in the RE compounds, in principle, originates from the CEF splitting of the ground *J*-multiplet for the open-shell 4*f* electrons, this can be best studied by inelastic neutron scattering experiment. Figure 9(a) provides inelastic neutron scattering spectrum obtained from FM $Au_{65}Ga_{21}Dy_{14}$ 1/1 AC, wherein a single flat mode at 3.3 meV can be distinguished. The integrated intensities at 2–5 meV are plotted as a function of the wavevector in Fig. 9(b). The wavevector dependence well follows that expected from the magnetic form factor of $Dy^{3+}$, suggesting that the flat mode corresponds to crystal field excitations of $Dy^{3+}$. Fig. 9(c) provides the energy transfer dependence of the integrated intensities at 1–2 $Å^{-1}$ measured at several temperatures. As the temperature increases, the peak position shifts to the lower energy transfer. The temperature dependence is quite close to that observed in the Tb-based FM 1/1 AC [15], suggesting the similarity in their CEF Hamiltonian. No significant signal was observed in the high energy regions, except for a weak flat mode observed at 170 meV (see Fig. S4 for details).

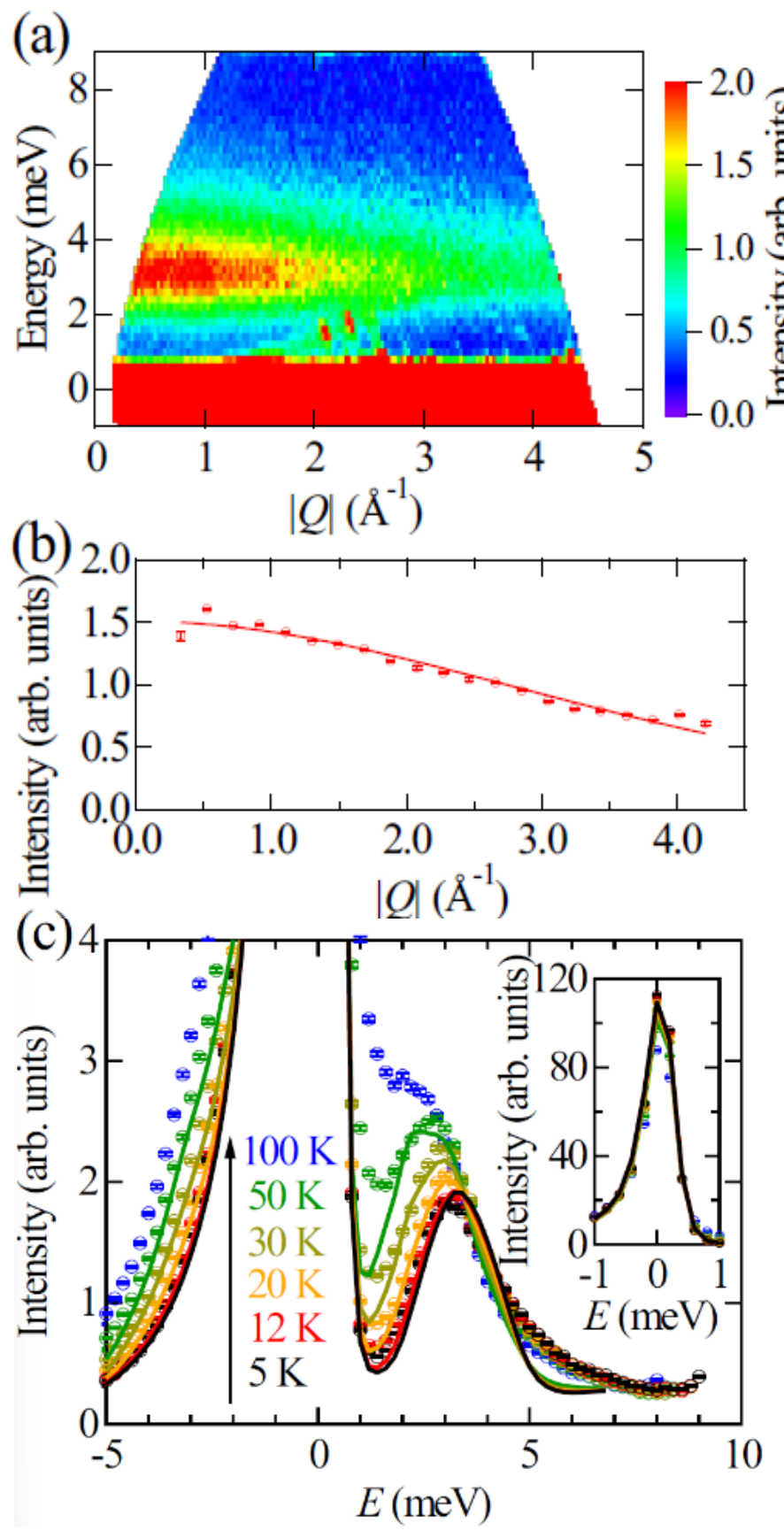

**Fig. 9.** (a) Inelastic neutron scattering spectrum of $Au_{65}Ga_{21}Dy_{14}$ collected at 6 K. Incident neutron energy is set to 12.74 meV. (b) Wavevector dependencies of the intensities integrated at the energy region of 2-5 meV. A red solid curve indicates the fit to the magnetic form factor of $Dy^{3+}$. (c) Energy dependencies of the intensities integrated at the wavevector region of 1-2 $Å^{-1}$. Solid curves indicate the fits to the point charge model (see text for details).

To estimate the CEF Hamiltonian, the excitation spectrum was analyzed by applying the point charge model used in the previous study [15]. Usually, the CEF excitations are analyzed using the operator equivalent technique [30]: the CEF Hamiltonian is expanded in terms of the Stevens' operator equivalents $O_k^q$ ($k = 2, 4, 6$ and $q = -k, -k+1, \cdots, k-1, k$), and the peak position and intensities are calculated from the eigenvalues and the matrix elements between their eigenvectors, respectively. The number of adjustable parameters is determined by the local symmetry of the magnetic ions. In the 1/1 AC such as $Au_{70}Si_{17}Tb_{13}$ [15] and $Au_{65}Ga_{21}Dy_{14}$, 15 parameters become adjustable due to a single mirror symmetry on the 24*g* site of the RE ions. In addition, ideally, the variance of the CEF potential induced by the site mixing present at nonmagnetic sites should be taken into account. The fit using all allowed parameters is, however, too underconstrained to obtain reasonable parameters from the observed single broad CEF mode. To overcome these difficulties, the CEF Hamiltonian was estimated by applying the point-charge model to the $Au_{68}Ga_{18}Dy_{14}$. The point-charge model allows one to decrease the number of adjustable parameters down to 3, and to introduce the variance of the CEF potential by calculating the statistical distribution of all the possible atomic configurations of the fractionally occupied sites. Although the point-charge model is not realistic for intermetallic compounds, it is quite helpful to model anisotropy of magnetic moments carried by RE ions. The point charge model yields the electrostatic potential as follows:

$$V_{\mathrm{CEF}} = \sum_{k=0,2,4,6} \sum_{q=-k}^{k} \langle r^k \rangle q_{kq} c_q^{(k)}(\theta,\phi)$$
$$q_{kq} \equiv \sqrt{\frac{4\pi}{2k+1}} \sum_i \frac{q_i e^2}{R_i^{k+1}} Y_q^{(k)*}(\theta,\phi) \qquad (1)$$
$$c_q^{(k)}(\theta,\phi) = \sqrt{\frac{4\pi}{2k+1}} Y_q^{(k)*}(\theta,\phi)$$

where e, $q_i$, and $(R_i, \theta_i, \phi_i)$ represent an electron charge, the charge of the *i*-th atom in the unit of *e*, and the position of the *i*-th atom relative to the RE ion in the spherical coordinate, respectively. The expectation values $\langle r^n \rangle$ for the 4*f* wavefunction in $Dy^{3+}$ ions are estimated as $\langle r^2 \rangle = 0.2188$ $Å^2$, $\langle r^4 \rangle = 0.1180$ $Å^4$, and $\langle r^6 \rangle = 0.1328$ $Å^6$ from the Dirac–Fock calculation [31]. The matrix elements of the spherical tensor operator $c_q^{(k)}$ are calculated as:

$$\langle J, J_{z1} \left| c_m^{(k)} \right| J, J_{z2} \rangle = (-1)^{J-J_{z1}} \times \begin{pmatrix} J & k & J \\ -J_{z1} & q & J_{z2} \end{pmatrix} \langle J \| c^{(k)} \| J \rangle \quad (2)$$

where $\langle J \| c^{(k)} \| J \rangle$ ($k$ = 2, 4, 6) corresponds to the reduced matrix element for $Dy^{3+}$ ions, $-4\sqrt{17/105}$, $-16/33 \cdot \sqrt{323/91}$, and $20/143 \sqrt{1615/39}$. The differential cross-section of the CEF excitations was calculated by taking the statistical average,

$$\left( \frac{d^2\Omega}{d\sigma dE} \right) = C \sum_{J_{z1}, J_{z2}} \sum_{\mu=x,y,z} P_{J,J_{z2}} \left| \langle J, J_{z1} | J_\mu^2 | J, J_{z2} \rangle \right|^2 \times \delta\left(\omega - E_{J,J_{z1}} + E_{J,J_{z2}}\right), \qquad (3)$$

under five thousand different atomic configurations. Here, $C$ is a constant factor including the magnetic form factor (the $Q$ dependence is ignored since it is unnecessary to fit the integrated intensities collected at the same wavevector range) and $P_{J,J_{z2}}$ represents the occupation of the eigenstate, $|J, J_{z2}\rangle$.

The energy transfer dependence is simulated using the point charge model based on the atomic structural of the $Au_{68}Ga_{18}Dy_{14}$, whose composition is quite close to the measured compound. The model includes three point-charge adjustable parameters $q_{Au}$, $q_{Ga}$ and $q_{Dy}$. The point charges for one element species (Au, Ga or Dy) are assumed to be the same regardless of their site symmetry. The fractional sites, Au4/Ga4 and Ga6/Au6 and the cluster center sites, within the radius $R_{in} < 8$ Å from the center $Dy^{3+}$ atom, is considered to statistically occupied either by Au or Ga (with the probability given by the occupancy). To reduce the computational cost, occupancy-weighted averages of Au and Ga valences are used for the fractional sites at $R_{in} < R < 30$Å. The resulting optimal point-charge parameters are $q_{\mathrm{Au}} = 0.096$, $q_{\mathrm{Ga}} = 0.55$, and $q_{\mathrm{Dy}} = -0.11$. The calculated inelastic spectra with the optimal charge parameters are shown by the solid curves in Fig. 9(c). Clearly, in the paramagnetic temperature range up to 50 K, the inelastic neutron scattering spectra is reasonably reproduced by the point-charge model with the statistically distributed Au and Ga atoms.

It is worth noting that the Ising character of the magnetic moments can be explained by the point-charge model as follows. The dominant terms in the CEF Hamiltonian (1), which are estimated by taking an average of five thousand atomic configurations, are $q_{20} = -0.009$ and $q_{22} = -0.005$. The principal axis is defined along the moment direction as described in the next paragraph. The dominant negative $q_{20}$ term suggests the Ising-anisotropy of Dy atoms induced by the second-order uniaxial CEF term $H_0 = -DJ_z^2$: eigenfunctions of the CEF Hamiltonian are mainly contributed by $|J_z\rangle$ ($J_z = -15/2, -13/2, \ldots\ldots, 15/2$) with the ground state ~|+-15/2>. Then, the differential cross section (3) is dominated by the transition between ($n$, $n$+1)-th and ($n$+2, $n$+3)-th CEF levels, whose energy decreases with increasing $n$ [5, 10]. Fig. 10 shows the CEF levels under five thousand randomly distributed atomic configurations. The inelastic neutron scattering spectrum reflects the transition between the ground doublet and the lowest excited doublet at low temperatures, while the transitions among ($n$, $n$+$1$)-th and ($n$+2, $n$+3)-th states ($n$ = 3, 5, ……, 13) become prominent at high temperatures. As a result, the strong Ising-anisotropy results in the appearance of a single peak and the peak shift with increasing temperature, as observed in Fig. 9(c).

Further, from the optimal point-charge parameters, the principal axes of the magnetic moment distribution are estimated by diagonalizing the expectation value matrix,

$$\begin{pmatrix} \langle J_x^2\rangle & \langle J_xJ_y\rangle & \langle J_xJ_z\rangle \\ \langle J_yJ_x\rangle & \langle J_y^2\rangle & \langle J_xJ_y\rangle \\ \langle J_zJ_x\rangle & \langle J_zJ_y\rangle & \langle J_z^2\rangle \end{pmatrix} \tag{4}$$

The eigenvector with the largest eigenvalue should represent the moment direction. The moment direction estimated by the averaged potential (setting $R_{\mathrm{in}}$ as 0) is $\langle J\rangle \parallel (0.00, 0.77, -0.64)$, while the moment direction estimated from the statistical average is $\langle J\rangle \parallel (0.00, 0.86, -0.52)$. This direction is 15 and 34 degrees away from to the magnetic moment direction of (0.00, 5.35, −1.59) of $Au_{68}Ga_{18}Dy_{14}$ 1/1 AC (AFM) and (2.52, 4.52, −0.92) of $Au_{60}Ga_{26}Dy_{14}$ 1/1 AC (FM), respectively. The average value of $g_J\sqrt{\langle J_z^2\rangle}$ along the principal axis is estimated as ~ 5.5 and ~ 5.1. Note that the moment direction becomes (1.00, 0.00, 0.00) if the point charge model with $q_{\mathrm{Au}} = 0.10$, $q_{\mathrm{Ga}} = 0.55$, and $q_{\mathrm{Dy}} = -0.11$ is applied to the structure of $Au_{60}Ga_{26}Dy_{14}$. The large difference in the magnetic anisotropy is induced by the change in the local structure around Dy atoms, as we discussed in Section 2.1. Given sensitivity of the point charge model to the atomic positions of Au and Ga atoms, the difference in the moment direction of AFM and FM structures may be also explained by the change in the local environment around Dy atoms.

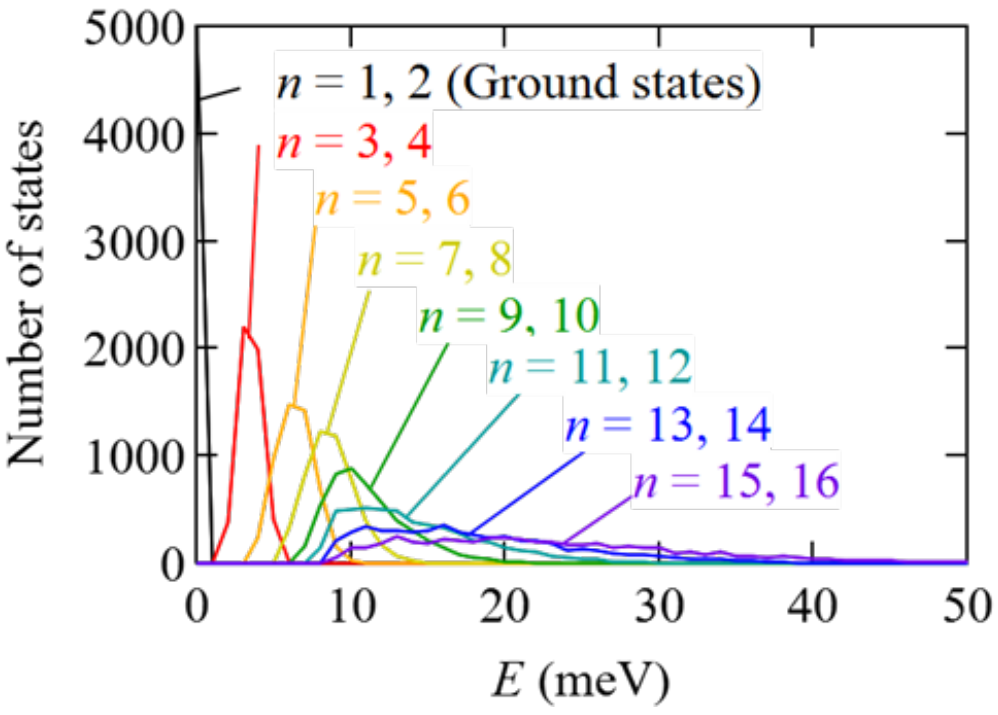

**Fig. 10.** Distribution of Dy CEF levels ($n$ = 1, 2, …, 16) under five thousand randomly distributed atomic configurations. All the CEF levels are doubly degenerated.

### III. II. iv. Magnetic phase diagram

With the information about magnetic structures of the whirling FM and AFM 1/1 ACs in hand, it is crucial to determine microscopic parameters contributing to stabilization of these structures. One of the important parameters in that regard is the ratio of next-nearest to nearest interactions ($J_2/J_1$). This parameter has been shown to play a prominent role in realizing magnetic states in non-Heisenberg Tsai-type compounds from a theoretical [18] viewpoints. To estimate $J_2/J_1$ for the present whirling AFM structure, one approach is to reproduce the field-dependent magnetization through calculations, focusing on the consistency in the position of the field-induced meta-magnetic transition. For this purpose, a simple icosahedron model composed of 12 Ising spins and Heisenberg interactions is considered. The interaction strength can be estimated from the following Hamiltonian:

$$H = -J_1 \textstyle\sum_{NN} S_i \cdot S_j - J_2 \sum_{NNN} S_i \cdot S_j + g\mu_B H \sum_i S_i \,. \tag{5}$$

wherein, the first and second terms correspond to the nearest and next-nearest neighbor interactions, respectively, while the third term accounts for the Zeeman interaction with an external magnetic field.

Figure 11 provides a series of calculated magnetization curves with $J_1$ and $J_2$ ranging from 0.006 to 0.015 and 0.066 to 0.070, respectively (shown by dashed lines) alongside the experimental

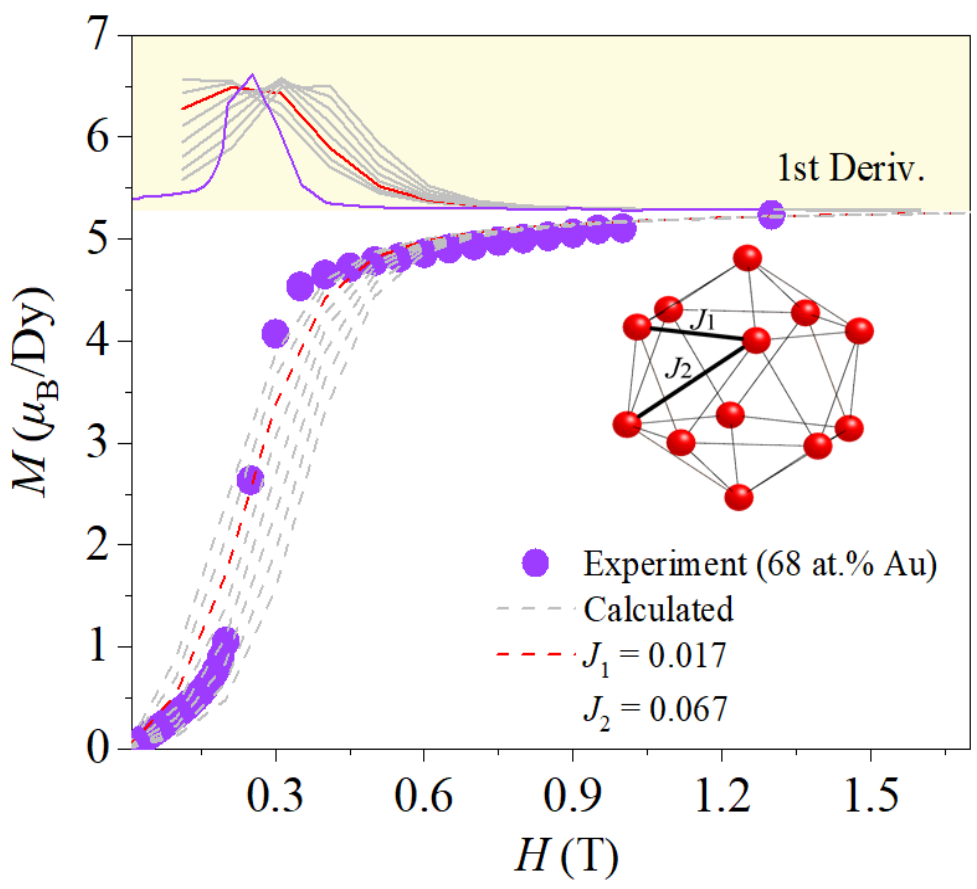

**Fig. 11.** An experimental field ($H$) dependence magnetizations curve of AFM $Au_{68}Ga_{18}Dy_{14}$ 1/1 AC superimposed with a series of calculated magnetizations curves within a field range of $0 < H < 1.7$ T with a first derivative curves provided on the top. A red dashed line corresponding to $J_1 = 0.017$ and $J_2 = 0.067$ best reproduced the experimental curve.

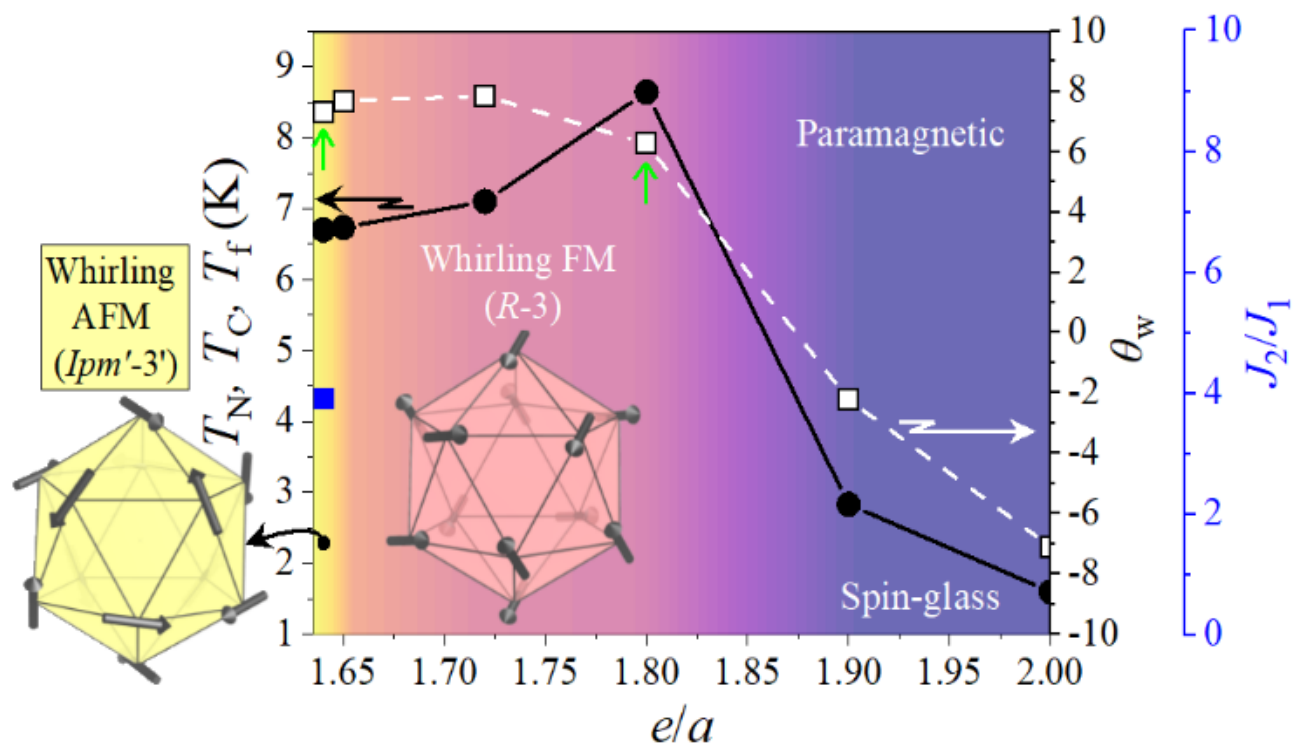

**Fig. 12.** A magnetic phase diagram of the Au-Ga-Dy 1/1 ACs showing *e/a* dependence of $T_C$, $T_N$ or $T_f$ (black markers) and $\theta_w$ (white markers). The yellow, red, and navy-blue background colors represent whirling AFM, whirling FM, and spin-glass regimes within the *e/a* parameter space, respectively. A blue marker represents an estimated $J_2/J_1$ of the AFM $Au_{68}Ga_{18}Dy_{14}$ 1/1 AC based on the calculation results employing Heisenberg Hamiltonian in equation (5). The corresponding magnetic structures of the whirling AFM and FM orders are also provided.

*M–H* curve of the $Au_{68}Ga_{18}Dy_{14}$ 1/1 AC (displayed with filled markers) within a field range of $0 < H < 1.7$ T. A first derivative of the curves is provided on the top of the figure. A qualitative comparison of the position of meta-magnetic transitions in the experimental and calculated curves reveals an agreement when $J_1 = 0.017$ and $J_2 = 0.067$ (colored dashed line in Fig. 11). By employing $\theta_w = 5J(J+1)(J_1 + J_2)/3$ ($J = 15/2$), these $J_1$ and $J_2$ values further yielded $\theta_w$ = 8.9 K which is close to $\theta_w$ = 7.3 K derived experimentally from $Au_{68}Ga_{18}Dy_{14}$ 1/1 AC. The resultant $J_2/J_1 = 3.9$ is further consistent with a range of $2 < J_2/J_1 < 4$ predicted theoretically from calculation for the whirling AFM phase in Dy-contained 1/1 AC [18].

Lastly, based on the results derived from both bulk magnetization and neutron experiments, a comprehensive magnetic phase diagram is developed for the Dy-contained 1/1 ACs within the *e/a* parameter space, as shown in Fig. 12. In the Figure, the left vertical axis represents a transition temperature [either Néel ($T_N$), Curie ($T_C$) or freezing ($T_f$)], while the right vertical axes correspond to the $\theta_w$ and the $J_2/J_1$. In the figure, the two samples indicated by green arrows correspond to $Au_{60}Ga_{26}Dy_{14}$ and $Au_{68}Ga_{18}Dy_{14}$ that are studied by PND. Rest of the data points in Fig. 12 are extracted from magnetic susceptibility and inverse susceptibility results, provided separately in Fig. S5 of Supplementary Material.

The developed phase diagram is quite informative from various viewpoints. First, contradictory to common expectation of negative $\theta_w$ from traditional AFM phases, the whirling AFM order is stabilized by positive $\theta_w$. This is because strong Ising-anisotropy can make the noncoplanar AFM structure stable under dominant FM interactions. Note that the magnitudes of the transition temperature and the magnetic moments in the present Au-Ga-Dy 1/1 ACs are smaller than those in the Au-Ga-Tb counterparts [10,13], despite the de Gennes parameter of Dy being larger than Tb. This might be due to the randomness in the CEF potential induced by the random occupation of Au and Ga atoms around Dy atoms. However, both the Au-Ga-Tb and Au-Ga-Dy 1/1 ACs include mixed ligand sites in a similar manner, and no apparent difference has been realized between the crystal structures of both compounds. The other possibility is the difference induced by the RE atoms. Note that all the CEF levels possess Kramers degeneracy for $Dy^{3+}$ ions. Electric quadrupole interactions, which can be only present in non-Kramers ions [32,33], may contribute to enhance the transition temperature.

## VI. CONCLUSION

Comprehensive experiments and analyses are performed in the non-Heisenberg Tsai-type 1/1 approximant crystals (ACs) in the Au-Ga-Dy system using magnetic susceptibility, neutron diffraction, and inelastic neutron scattering techniques. The results unveiled noncoplanar "*whirling*" ferromagnetic and antiferromagnetic orders, structurally analogous to those in Tb- and Ho-containing counterparts. Through detailed comparison of the crystal electric field excitations and spin configurations in Tb-, Dy- and Ho-containing ACs, a universal mechanism that stabilizes whirling orders in non-Heisenberg ACs is established; ferromagnetic intra-cluster interactions and strong Ising-like anisotropy.

## ACKNOWLEDGMENT

This The authors acknowledge Akiko Takeda for assistance in the synthesis of the materials. This work was supported by Japan Society for the Promotion of Science through Grants- in-Aid for Scientific Research (Grants No. JP19H05817, JP19H05818, JP19H05819, JP21H01044, JP22H00101, JP22H04582, 23KK0051 and No. JP24K17016), Murata Science and Education Foundation, Japan Science and Technology agency, CREST, Japan, through Grant No. JPMJCR22O3. The experiments at JRR-3 and at the Materials and Life Science Experimental Facility of the J-PARC were supported by the General User Program for Neutron Scattering Experiments (No. 22511) of Institute for Solid State Physics, University of Tokyo and a user program (Proposal No. 2020L0300), respectively. The authors acknowledge Dr. Kazuyasu Tokiwa from Tokyo University of Science for his assistance in performing magnetic susceptibility measurements on some of the samples studied.

## RERERENCES